\documentstyle[multicol,aps,prl,epsf]{revtex}

\begin{document}
\pagestyle{empty}

\newcommand{\bc}{\begin{center}}
\newcommand{\ec}{\end{center}}
\newcommand{\be}{\begin{equation}}
\newcommand{\ee}{\end{equation}}
\newcommand{\beqn}{\begin{eqnarray}}
\newcommand{\eeqn}{\end{eqnarray}}

\begin{multicols}{2}
\narrowtext
\parskip=0cm

\noindent
{\large\bf Comment on ``General Method to Determine Replica Symmetry Breaking
Transitions''}
\smallskip

In a recent letter Marinari et al.~\cite{marinari} introduced a new method
to study spin glass transitions and argued
that by probing replica
symmetry (RS) as opposed to time reversal symmetry (TRS), their method
unambiguously shows that replica symmetry breaking (RSB) occurs in
short-range spin glasses. In this comment we show that while the method
introduced in \cite{marinari} is indeed useful for studying transitions
in systems where TRS is absent (such as the p-spin model studied by them), 
the conclusion that it shows the 
existence of RSB in short-range spin glasses is wrong.

The analysis of Marinari et al. is
based on a new quantity which for systems with TRS is given by
\begin{equation}
G(T, L) = {{\left[\langle q^2 \rangle^2\right] - 
\left[\langle q^2 \rangle\right]^2}
\over{
\left[\langle q^4 \rangle\right] - 
\left[\langle q^2 \rangle\right]^2}}
\end{equation}
where $\langle ..\rangle$ and $\left[ .. \right]$ denote thermal and disorder
averages respectively,
$T$ is the temperature, 
$L$ is the system size, and $q$ is
the standard overlap between two replicas. 
For the Ising spin glass without a field in
four dimensions, 
Marinari et al. found 
that $G(T,L)$ exhibits two distinct kinds of
behavior quite clearly: a high temperature phase where $G(T,L)$ decreases
with system size as $1/{L^d}$ in $d$ dimensions
and a low temperature phase where $G(T,L)$
increases with system size and saturates at a constant value 
close to $1/3$.
They interpreted this as indicating the existence of RSB
in these systems (They also studied the Ising spin glass in a field
which we will discuss towards the end of this comment.). 

However, within a droplet like picture 
it is not clear
how $G(T,L)$ behaves as $L\to\infty$ because 
$P(q) \to {1\over 2}\left[\delta(q-q_{EA})+\delta(q+q_{EA})\right]$ and 
both the numerator
and the denominator are zero in this limit. 
Motivated by this we studied $G(T,L)$ for the three-dimensional
Ising spin glass within the Migdal-Kadanoff approximation used 
recently by Moore, Bokil and Drossel\cite{us}.
These authors showed that even though the asymptotic behavior is
described by the droplet theory, finite systems can exhibit many
of the features associated with RSB. 
In Fig.1 we show $G(T,L)$ as 
a function of $T$ for a range of system sizes $L = 4, 8, 16$, the
averages being performed over $20000-50000$ samples.
Just as in the simulation reported in \cite{marinari} we find two distinct
regimes, a high temperature phase with $G(T,L)$ decreasing with
sytem size and a low temperature phase where $G(T,L)$ increases with
system size saturating at a value around $1/3$. Clearly then, this behavior
of $G(T,L)$ cannot be interpreted as evidence of RSB. What is
happening here is that finite systems show non-self-averaging of 
the overlap distribution function (and a non zero limit for
$G(T,\infty)$ in the low temperature phase) but there is self-averaging
in the infinite system limit where one has a droplet-like 
$P(q)$. The mistake in Ref.\cite{marinari} was to
note that the numerator in Eq.(1) would vanish in the thermodynamic
limit with RS, while overlooking the fact that the denominator would vanish
as well. Thus, $G(T,L)$ can have a non zero limit.
As an aside we note that even for the rather trivial case of a 
single random bond (connecting two spins) 
it is possible to 
prove that
$G = 1/3$ at $T=0$ provided
that the bond distribution function has a non-zero weight at the origin.

\begin{figure}
\centerline{
\epsfysize=0.7\columnwidth{\epsfbox{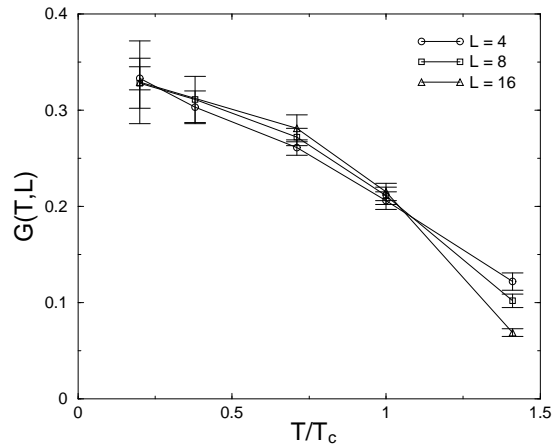}}}
\narrowtext{\caption{G(T, L) in the Migdal-Kadanoff approximation for 
L = 4, 8, 16}}
\label{fig1}
\end{figure}

Thus we have demonstrated that the data reported in \cite{marinari}
for the Ising spin glass without a field do not give any evidence for
RSB in this system. If the authors of \cite{marinari} had found
convincing evidence for a transition in the case with a field, that
would indeed have been evidence for RSB. But we believe that their data 
for that case 
do not allow any conclusive statement to be made (there is no
apparent crossing of the curves nor is it clear that there are
two distinct kinds of behavior corresponding to high and low temperatures).
In summary, while the method developed in \cite{marinari} is
useful for some problems, it does not give evidence for RSB
in Ising spin glasses.

\bigskip
\noindent
Hemant Bokil, A. J. Bray, Barbara Drossel, and M. A. Moore

{\small

Department of Physics and Astronomy

University of Manchester

Manchester M13 9PL, U.K.

}
\bigskip
\noindent
Date: 29 September, 1998


\end{multicols}

\end{document}